\documentclass{PoS}

\usepackage[mathletters]{ucs}   
\usepackage[utf8x]{inputenc}
\usepackage{xspace}
\usepackage{siunitx}
\usepackage{amssymb}
\usepackage{booktabs}
\usepackage{subfig}
\usepackage[percent]{overpic}
\usepackage{lineno}

\newcommand{\hess}{H.E.S.S.\xspace}
\newcommand{\ct}[1]{CT\,#1\xspace}
\newcommand{\vhe}{VHE\xspace}
\newcommand{\tmva}{TMVA\xspace}
\newcommand{\cog}{c.o.g.\xspace}

\newcommand{\fref}[1]{Fig.~(\ref{fig:#1})} 
\newcommand{\tref}[1]{Tab.~(\ref{tab:#1})} 

\DeclareSIUnit\pe{p.e.}
\title{A Neural Network-Based Monoscopic Reconstruction Algorithm for \hess~II}

\ShortTitle{Monoscopic Reconstruction for \hess~II}

\author{\speaker{Thomas Murach}\\
        Humboldt-Universität zu Berlin\\
        E-mail: \email{murach@physik.hu-berlin.de}}

\author{Michael Gajdus\\
       Humboldt-Universität zu Berlin\\
       E-mail: \email{mgajdus@physik.hu-berlin.de}}
\author{Robert Daniel Parsons\\
       Max-Planck-Institut für Kernphysik\\
       E-mail: \email{Daniel.Parsons@mpi-hd.mpg.de}}

\abstract{The \hess experiment entered its phase~II with the addition of
a new, large telescope named \ct{5} that was added to the centre of the
existing array of four smaller telescopes. The new telescope is able to detect
fainter air showers due to its larger mirror area, thereby lowering the
energy threshold of the array from a few hundred \SI{}{GeV} down to
$\mathcal{O}(\SI{50}{GeV})$. Due to
the power-law decrease of typical γ-ray and cosmic-ray spectra of astrophysical
sources a majority of
detected air showers are of low energies, thus being detected by \ct{5} only,
which motivates the need for a reconstruction algorithm based on information from
\ct{5} alone. By exploiting such monoscopic events
the \hess experiment in phase~II becomes sensitive in an energy range not
covered by \hess~I and in which the Fermi LAT runs out of statistics.
Furthermore the chance of detecting transient phenomena like γ-ray bursts is increased
significantly due to the large effective area of \ct{5} at low energies.

In this contribution a newly developed reconstruction algorithm for monoscopic
events based on neural networks is presented. This algorithm uses multilayer perceptrons to
reconstruct the direction and energy of the particle initiating the air shower
and also to discriminate between gamma rays and hadrons. The performance
of this algorithm is evaluated and compared to other existing reconstruction
algorithms. Furthermore results of first applications of the algorithm to
measured data are shown.}

\FullConference{The 34th International Cosmic Ray Conference,\\
		30 July- 6 August, 2015\\
		The Hague, The Netherlands}

\begin{document}

\section{Introduction}
Ground-based very high energy (\vhe, $\sim \SI{30}{GeV}$ -
\SI{100}{TeV}) γ-ray experiments have
established a new astronomical field over the last 20 years, with more than 100
sources discovered already \cite{tevcat}.
Current imaging atmospheric Cherenkov telescope
(IACT) experiments like MAGIC, VERITAS and \hess
are able to detect the faint Cherenkov light emitted in extended
air showers induced by \vhe particles interacting with nuclei in the Earth's
atmosphere. The \hess experiment \cite{hess} entered its phase~II in 2012 with the addition
of a fifth telescope named \ct{5} to the centre of the existing array of four
\SI{12}{\meter} telescopes. This new telescope is the largest of its kind, with
a mirror area of \SI{600}{\square\meter}, able to detect air showers from
low-energy particles. Therefore the lower end of the observable energy range can be
decreased to $\mathcal{O}(\SI{50}{GeV})$, compared to a few hundred GeV in case
of the \hess~I array. Due to the power-law decrease of the number of incident
particles as a function of their energy a majority of all events will be detected
by \ct{5} only. A newly developed reconstruction chain called \textit{MonoReco},
implementing a full event reconstruction using information from \ct{5} only, is
presented here. It is the first monoscopic reconstruction algorithm based on
multivariate analysis techniques used within the \hess experiment.

\section{MonoReco - A Monoscopic Reconstruction Algorithm}
Several classes of astrophysical objects are able to accelerate charged
particles up to {\vhe}s. Those particles, mainly protons, heavier nuclei,
electrons and positrons, are deflected by
interstellar and intergalactic magnetic fields. Thus, in the energy range
accessible to IACTs, they are isotropised on their way to Earth. To be able to
study the acceleration
processes occurring in individual sources γ rays produced by those charged particles
through e.g. inverse Compton scattering or $π^0$ decay can be analysed. For a
typical source the number
of γ rays detected by IACTs is lower than the number of detected hadrons by a factor
of $\mathcal{O}(1000)$. Therefore it is important to efficiently distinguish
between those particle classes as well as to reconstruct the particles' energies
and incident directions.

Most reconstruction concepts depend heavily on stereoscopy, i.e. information
from at least two telescopes, to be able to reconstruct all of the main properties
of the primary particles like their direction and energy
as well as the type of particle. This is no longer possible for low-energy
showers that are only seen with \ct{5}.
In order to extend the energy range accessible to \hess towards the regime in
which the Fermi Large Area Telescope (LAT) \cite{fermi} operates
($\lesssim \SI{100}{GeV}$) it is necessary to analyse monoscopic events.
Such a low-energy analysis is especially needed in
the case of transient events like γ-ray bursts or flaring activity from active
galactic nuclei because of the $\sim 10^4$ times larger effective area of \ct{5}
with respect to the effective area of the Fermi LAT.

The reconstruction chain presented here is capable of performing a full event
reconstruction. It makes use of the so-called
Hillas parameters \cite{hillasparams},
the moments of the distribution of the measured intensities in the individual
pixels of the camera of \ct{5}, and multivariate analysis techniques to
determine all relevant properties of the incident particle.

Before multivariate analysis techniques are applied, several preprocessing steps
are performed.
Noise is removed from the images by requiring two neighbouring pixels to contain
minimal intensities of 5 and \SI{10}{photo} electrons (\SI{}{\pe}) or vice versa.
After this image cleaning has been performed,
several cuts are applied to guarantee a sufficient intensity and number of pixels.
Also the nominal distance, i.e. the angular distance of
the center of gravity (\cog) of the Hillas ellipse from the centre
of the camera, is cut on to make sure the image is not truncated at the camera
edges. The cut values are summarised in \tref{cuts} for several cut configurations.

\begin{table}[tb]
  \centering
  \begin{tabular}{l|llll}
    \toprule
    {\bf Cut parameter} & & {\bf \textit{std}} & {\bf \textit{loose}} & {\bf \textit{extraloose}}\\
    \midrule
    Number of pixels & & $>3$ & $>3$ & $>3$\\
    Intensity & & $>\SI{60}{\pe}$ & $>\SI{60}{\pe}$ & $>\SI{35}{\pe}$ \\
    Nominal distance & & $<\SI{1.15}{\degree}$ & $<\SI{1.15}{\degree}$ & $<\SI{1.15}{\degree}$\\
    ζ & & $>0.9$ & $>0.78$ & $>0.55$\\
    $\Theta$ & & \SI{0.13}{\degree} & \SI{0.14}{\degree} & \SI{0.23}{\degree}\\
    \bottomrule
  \end{tabular}
  \caption{Cut values used in the three different cut configurations introduced
  in the main text.}
  \label{tab:cuts}
\end{table}

In each of the three main analysis steps neural networks are used to determine
properties of the primary particles. The implementation of
multilayer perceptrons (MLPs) within the \tmva framework
\cite{tmva} was used in this
analysis. Before such neural networks can be used in an analysis they have to
be trained. In the reconstruction chain presented here
γ-ray Monte Carlo simulations were used in the training stage.

In general the shape of a γ-ray air shower image in the focal plane of the
telescope mirror is elliptical. For geometrical reasons the direction of the
initial particle is a point assumed to be located along the major axis of the
ellipse which is parameterised
as a Hillas ellipse. Based on this assumption the direction is reconstructed
using the \cog together
with a displacement value δ calculated as the distance between the \cog and the
direction to be reconstructed.
When analysing data δ is retrieved from the
previously trained MLP networks working in regression mode based on the
following input parameters: the width and length of the Hillas ellipse, the total
intensity divided by the area of the Hillas ellipse (called \textit{density}), the logarithm of the image
amplitude, the skewness and the kurtosis of the intensity distribution in the
camera.
As only the absolute value of δ can be retrieved from the MLP network it is
unclear in which direction
along the major axis of the ellipse the δ value has to be applied. This
degeneracy can be resolved by using the asymmetry of the intensity distribution
in the camera parameterised by the skewness.

The reconstruction of the energy of the initial γ ray is also based on MLP networks
working in regression mode using the same input variables as the MLP networks
mentioned above.

Finally the type of the primary particle has to
be determined. In order to accomplish this task an MLP network operating in
classification mode is used. The MLP network is provided with a set of variables
(the width and length of the Hillas ellipse, the density, the skewness and kurtosis
as well as the length over the logarithm of the intensity)
from Monte Carlo
simulations in case of γ rays and with parameters taken from so-called OFF runs,
i.e. observations of regions of the sky containing no known sources of γ rays, in
case of background events. Based on such input events of known
signal or background class the MLP network is able to construct a function which
calculates an output parameter ζ based on the set of input variables. This value
is used to perform a cut in order to discriminate between γ
rays and hadrons. The definition of the cut value is subject to an optimisation
procedure. Several science cases differing in the assumed spectral index of γ-ray
differential energy spectra were accounted for by the definition of
several cut sets. A \textit{std} cut set is defined such that the
Q factor, defined as the signal efficiency divided by the square root of the background efficiency, of
a source with a spectral index of $-2$ is maximal by
adjusting the cut on ζ and the radius of the signal region $\Theta$ accordingly. A
\textit{loose} cut set is defined similarly but using an index of $-3$. An
\textit{extraloose} cut set designed for pulsar analyses is defined to
obtain maximal effective areas at low energies. Cut values for ζ and $\Theta$ are
given in \tref{cuts}.

All neural networks used in this reconstruction chain
were trained for several observation conditions, differing in zenith angle,
azimuth angle and source offset from the centre of the camera.
In analyses of real data the MLP networks trained for simulated
positions adjacent to the real observation position are used to perform an
interpolation.

\section{Performance of the Reconstruction Chain}
In the following the performance of the reconstruction chain will be evaluated.
All events which have triggered \ct{5} were analysed regardless of the
trigger information of other telescopes in order to demonstrate the performance
over the entire coverable energy range. For example when performing split
observations with only \ct{5} observing one source events at all energies need
to be reconstructed monoscopically.

\subsection{Direction Reconstruction Performance}
The direction reconstruction performance can be studied by means of an
angular resolution distribution as presented in \fref{angres} for three
different zenith angles. In this
figure the \SI{68}{\percent} containment radius $\mathrm{R}_{68}$ of the
distribution of reconstructed directions around the direction of the simulated
γ rays is shown as a function of their simulated energy.
\begin{figure}[ht]
    \centering
    \subfloat[Angular resolution\label{fig:angres}]{%
        \begin{overpic}[width=0.48\textwidth]{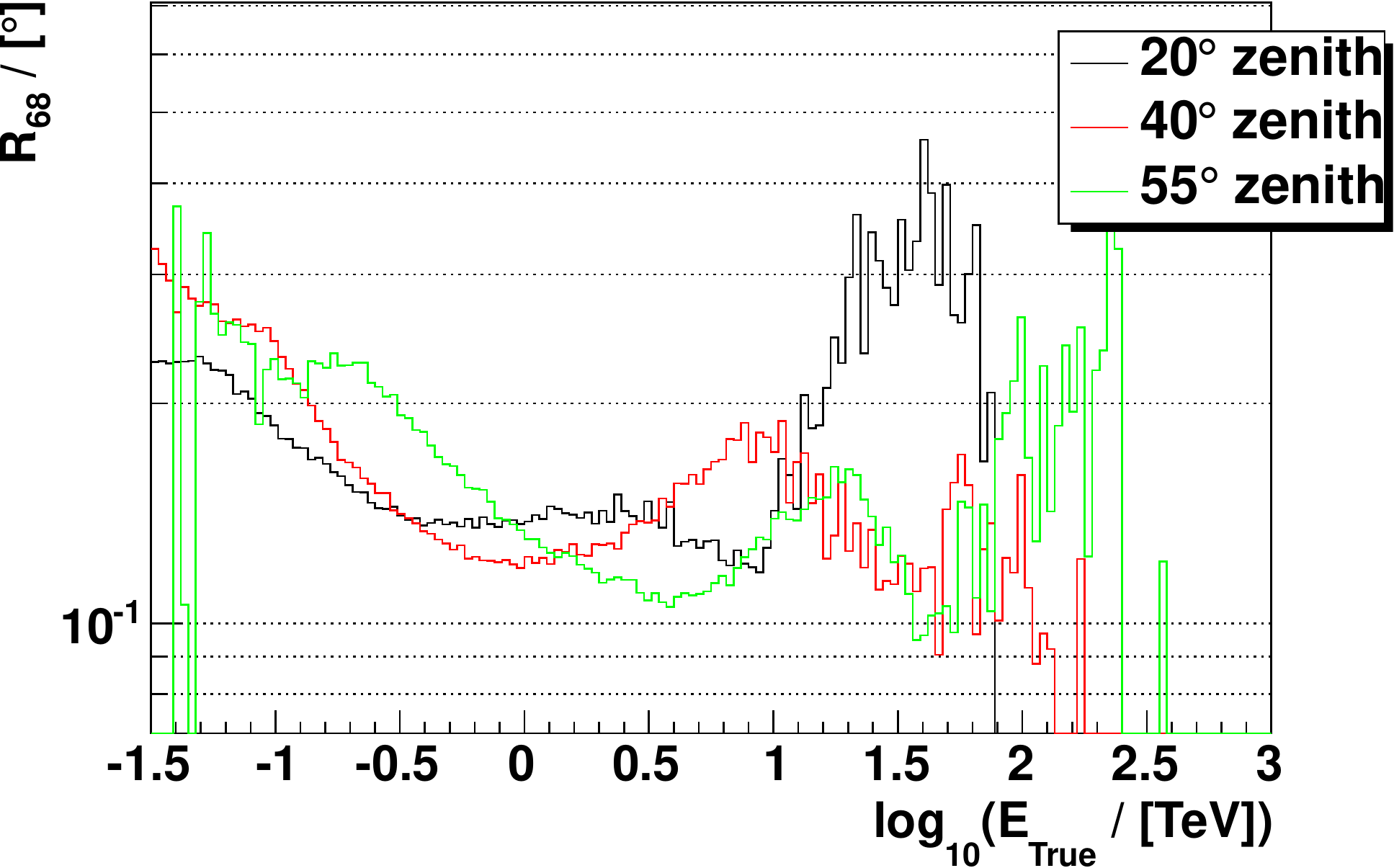}
        \end{overpic} }  \hfill
    \subfloat[Energy bias\label{fig:ebias}]{%
        \begin{overpic}[width=0.48\textwidth]{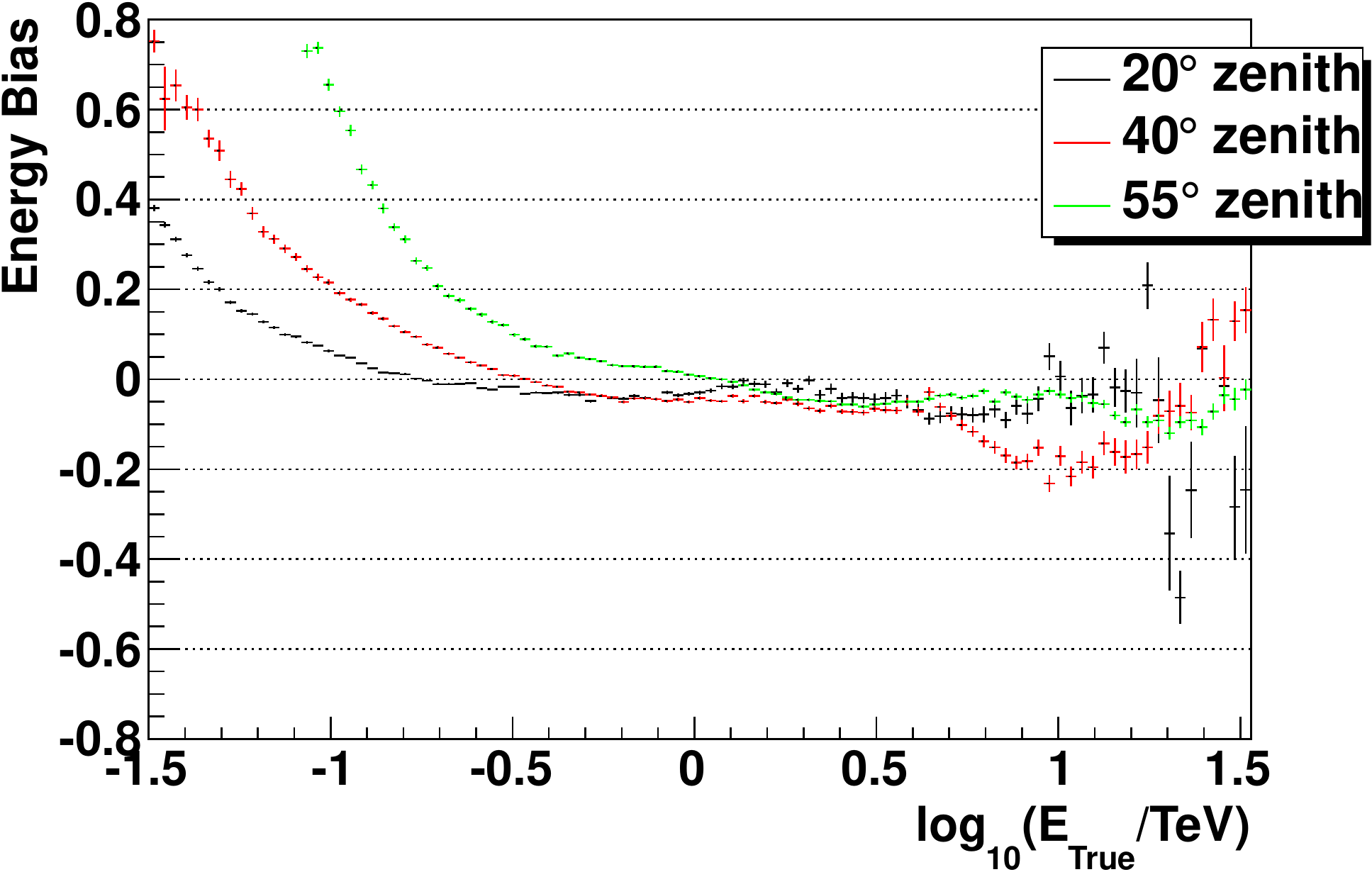}
        \end{overpic} }  \qquad
    \subfloat[Energy resolution\label{fig:eres}]{%
        \begin{overpic}[width=0.48\textwidth]{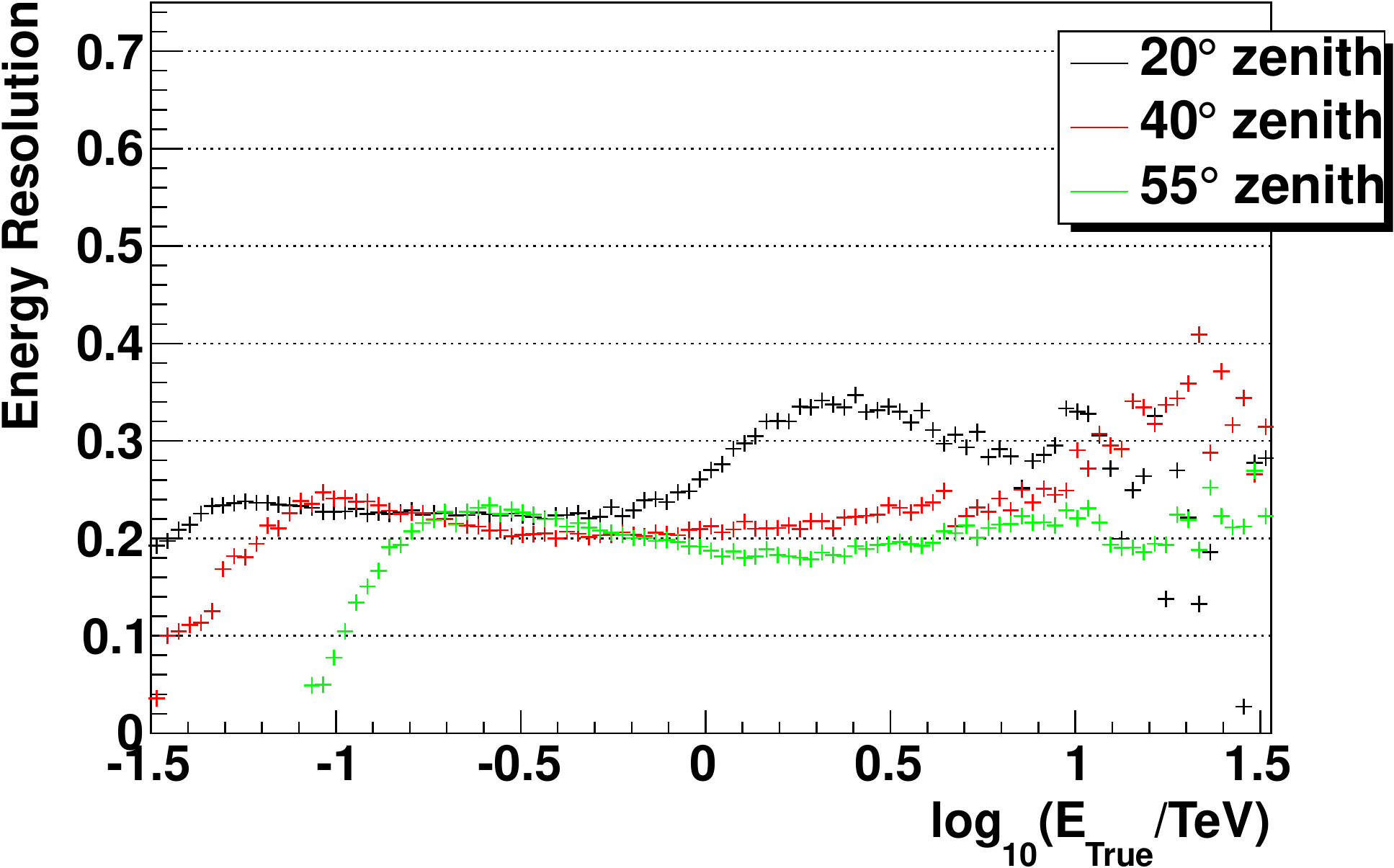}
        \end{overpic} }  \hfill
    \subfloat[Effective area\label{fig:effarea}]{%
        \begin{overpic}[width=0.48\textwidth]{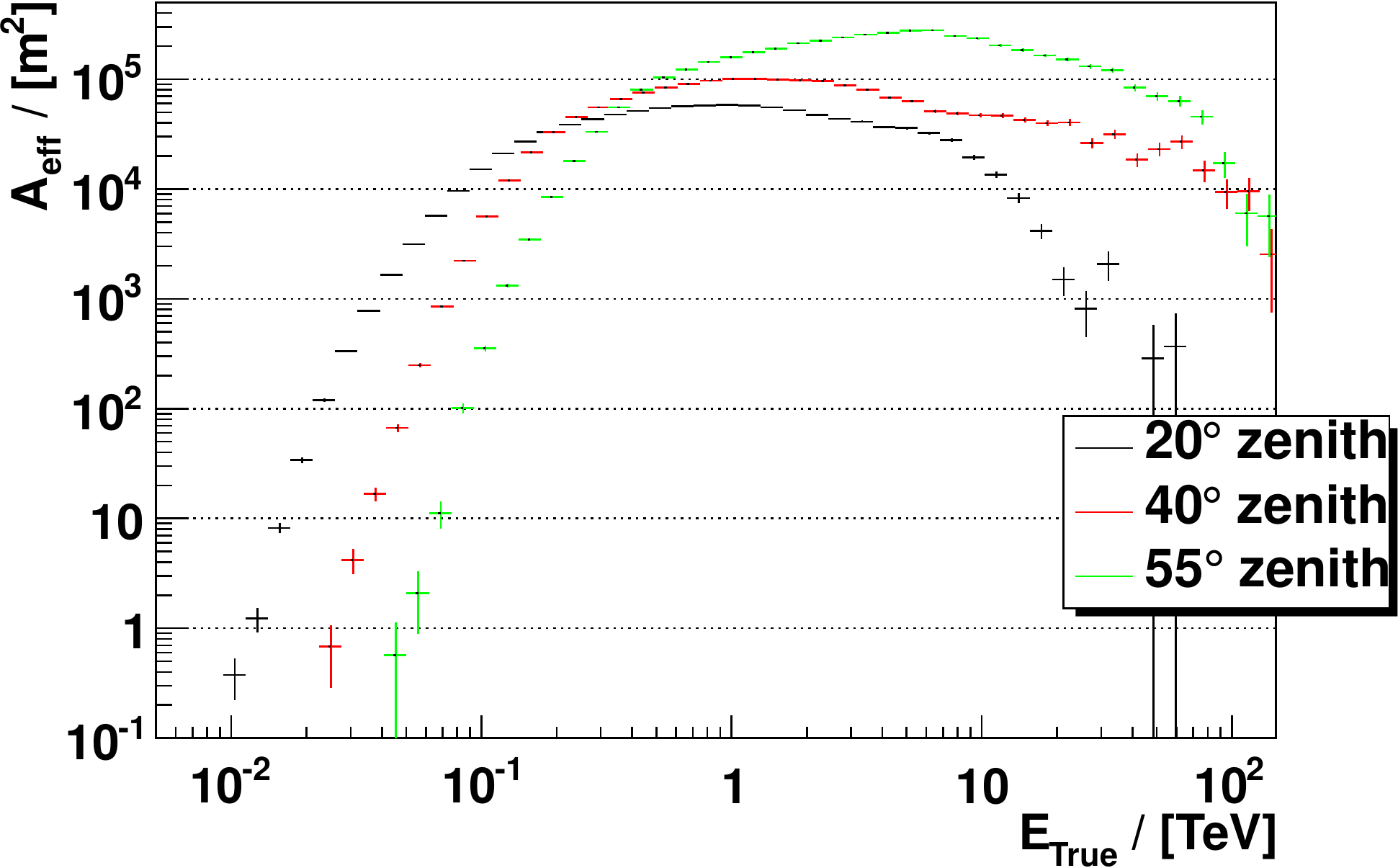}
        \end{overpic} }
  \caption{Instrument response functions corresponding to several zenith
          angles are shown in the above figures as functions of the
          true energy E$_{\mathrm{True}}$. An azimuth angle
          of \SI{0}{\degree} and a wobble offset of \SI{0.5}{\degree} was used
          to produce all curves. The displayed response functions
          are (a) the \SI{68}{\percent} containment radius R$_{68}$, (b)
          the energy bias, (c) the energy resolution and (d) the effective area.}
  \label{fig:irfs}
\end{figure}
The angular resolution at low energies is $\sim\SI{0.3}{\degree}$ and improves
with increasing energy as the shower images become brighter to reach a
minimum of $\sim\SI{0.1}{\degree}$, roughly comparable to values achieved by
the more advanced ImPACT analysis, which uses a template likelihood fit to determine γ-ray
properties \cite{impactarxiv}, \cite{impacticrc}. Towards highest energies the angular
resolution worsens as the shower images become so large that some images are
cropped at the camera edges. In general this leads to
poorer event reconstruction.

\subsection{Energy Reconstruction Performance}
The quality of the energy reconstruction can be quantified by the energy
reconstruction bias and resolution. The bias is the mean and the resolution is
the RMS of the distribution of the relative
deviation of the reconstructed energies from the true energy as a function of
the true energy. Both quantities can, by definition, only be calculated
from Monte Carlo data.

The bias curves corresponding to three different zenith angles are displayed in
\fref{ebias}. In all cases the bias is smaller than
\SI{10}{\percent} at energies up to $\sim\SI{5}{TeV}$.
At low energies the bias is positive because only γ-ray air showers
whose recorded intensities fluctuated upwards can be detected.

Energy resolution curves are shown in \fref{eres}. At the lowest
energies all shower images look approximately the same, thus the reconstructed
energies are very similar leading to a small resolution at lowest energies.
At higher energies the resolution stays almost constant around 20 -
\SI{30}{\percent}.
All values are very similar to the performance of the ImPACT analysis.

\subsection{Effective Area}
The effective area is defined as the γ-ray detection efficiency multiplied by
the area corresponding to the maximum impact distance of the simulated air
shower axes with respect to the telescope (\SI{500}{\meter} at \SI{20}{\degree}
zenith). Effective areas as a function of the true energy for
three different zenith angles are shown in \fref{effarea}.
At the lowest energies the detection efficiency is very low. As the showers get
brighter and more regular with increasing energy the efficiency increases
until a maximum is reached. At the highest energies the efficiency begins to
drop because the showers tend to get cropped at the camera edges leading to a
worse classification performance. The maximum value of the effective areas
increases with the zenith angle due to projection effects.

\subsection{Sensitivity}
The lowest point-source flux that can be detected with a significance of
5σ within \SI{50}{\hour} of observations in
an energy interval is called differential
sensitivity. This quantity expressed in units of the differential Crab nebula
flux is shown in \fref{sensitivity} as a function of the reconstructed energy.
\begin{figure}[t]
  \centering
  {%
  \begin{overpic}[width=0.79\textwidth]{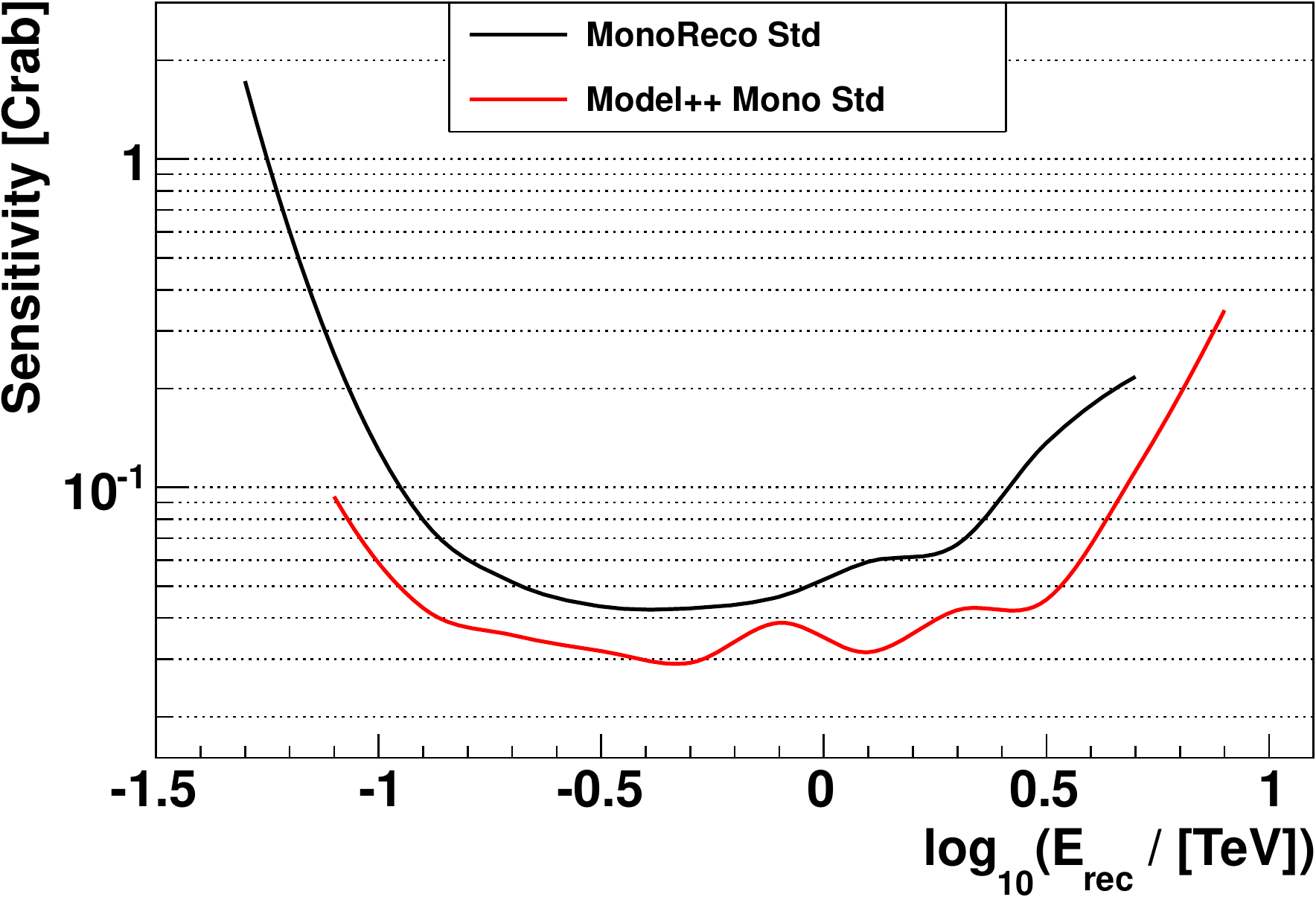}
    \put(29.5,50){\huge \textcolor{red}{\hess Preliminary}}
  \end{overpic} }
  \caption{Differential sensitivities obtained by the MonoReco (black)
           and Model++ (red, \cite{hesssens}) analyses in units of the
           differential Crab Nebula flux as
           a function of the reconstructed energy $\mathrm{E}_{\mathrm{rec}}$.
           The curves correspond to
           a zenith angle of \SI{20}{\degree}, an azimuth angle of
           \SI{180}{\degree} and a wobble offset of \SI{0.5}{\degree}.}
  \label{fig:sensitivity}
\end{figure}
The significances were estimated as $\textrm{N}_{γ}/\sqrt{\textrm{N}_{\textrm{Bkg}}}$,
with $\textrm{N}_γ$ being the number of excess events and $\textrm{N}_{\textrm{Bkg}}$
being the exposure-corrected number of background events. The
lowest detectable flux value per energy bin at a zenith angle of \SI{20}{\degree}
corresponds to \SI{4}{\percent} of the Crab Nebula flux. More advanced
reconstruction chains like ImPACT or Model++ \cite{hesssens}, \cite{modelpp}
reach values of $\sim\SI{3}{\percent}$, as shown in \fref{sensitivity}.

\section{Analysis of Crab Nebula Data}
The reconstruction chain presented in the previous sections was applied to
data from observations of the Crab Nebula. In total 16 observation intervals
lasting \SI{28}{\minute} each were analysed, resulting in a live time of
\SI{7.2}{\hour}. The observations were performed with a wobble offset of
\SI{0.5}{\degree} from the camera centre at a mean zenith angle of
\SI{47}{\degree}. Using the \textit{reflected background} method \cite{bgpaper}
4800 excess events were measured in the ON
region with respect to the number of events detected in OFF regions.
This corresponds to a significance of 91σ using Eq. (17)
from Li \& Ma \cite{lima}.

The $\Theta^2$ distributions shown on the left hand side of \fref{thetaSqrAndSignif}
visualise the distributions of the squared angular distances of the reconstructed shower directions
from the centres of the corresponding ON and OFF regions.
\begin{figure}[tb]
   \centering
   \subfloat{%
     \begin{overpic}[width=0.49\textwidth]{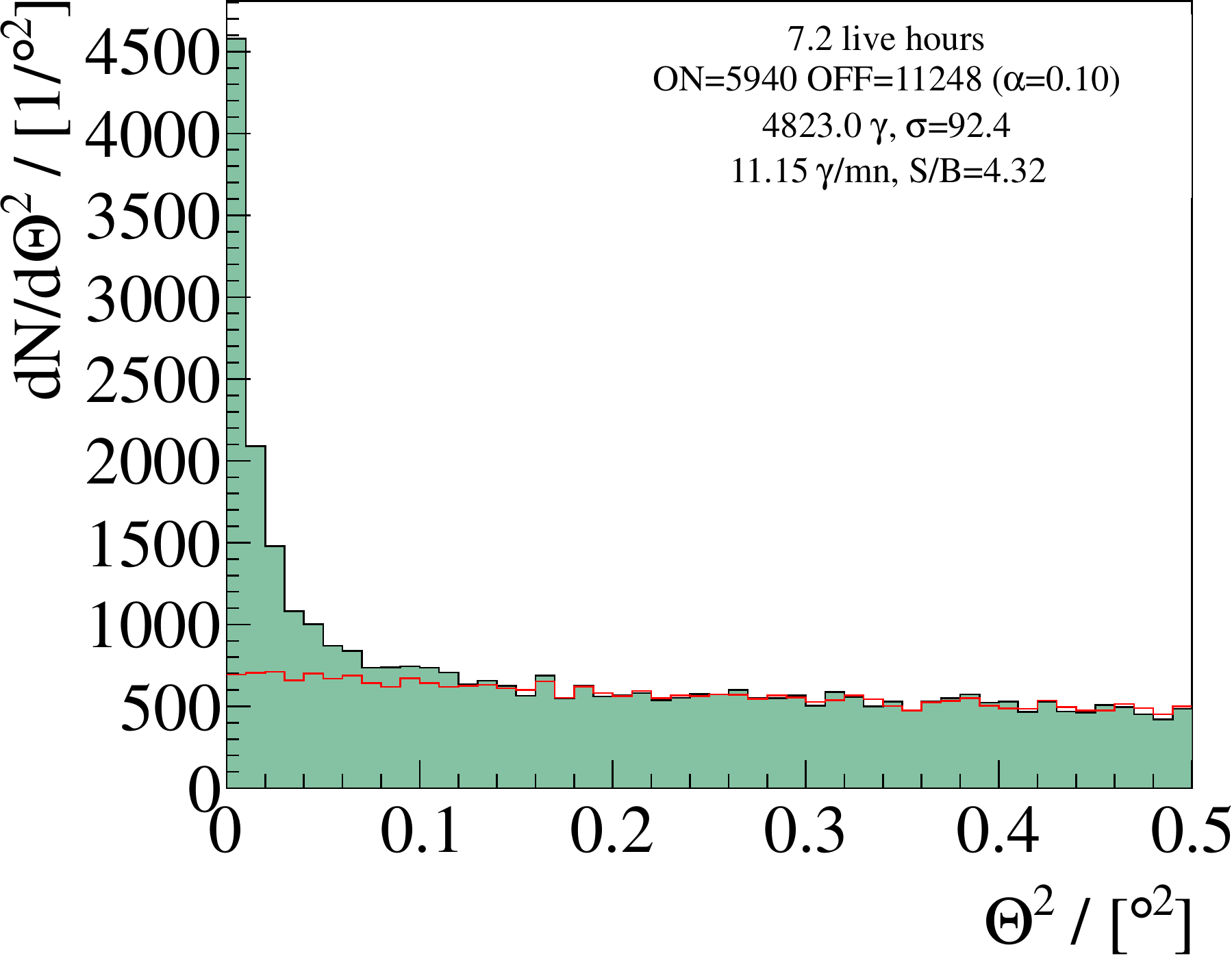}
       \put(45,26){\large \textcolor{red}{\hess Preliminary}}
     \end{overpic} }
   \subfloat{%
     \begin{overpic}[width=0.49\textwidth]{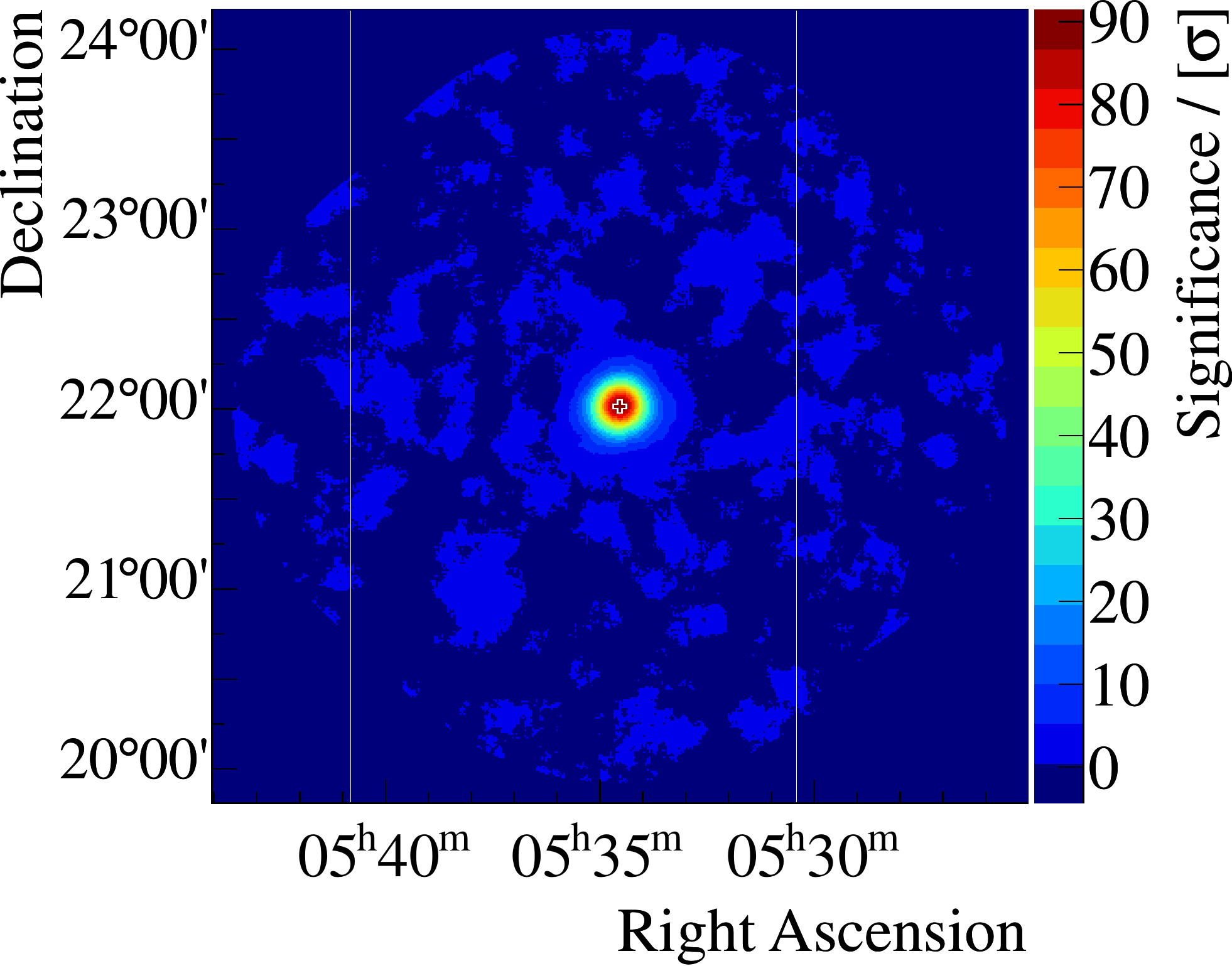}
       \put(35,26){\large \textcolor{red}{\hess Preliminary}}
     \end{overpic} }
   \caption{\textit{Left}: Squared angular distance of the reconstructed
           directions to the test positions in the centre of the ON (black) and
           OFF (red) regions. \textit{Right}: Sky map.}
   \label{fig:thetaSqrAndSignif}
\end{figure}
It can be seen that there is an excess towards the position of the Crab Nebula
whereas there is none towards the test positions in the OFF regions.

Significances can also be calculated from any position in the sky using the
\textit{ring background} method \cite{bgpaper}. The
resulting significance map is shown on the right hand side of
\fref{thetaSqrAndSignif}. It can be seen that there are no artefacts visible in
the field of view and that the source is point-like and centred on the position
determined from radio measurements conducted with the VLBI \cite{crabpos}.

At zenith angles around \SI{45}{\degree} the safe energy threshold is approximately
\SI{150}{GeV}, compared to a threshold of $\sim\SI{300}{GeV}$ in case of \hess~I.
The threshold is defined as the energy at which the effective area reaches
\SI{10}{\percent} of its maximum value. This definition does not take the shape
of the energy bias or energy resolution curves into account, thus the bias could,
in principle, be arbitrarily large. Therefore a forward-folding technique was
used to derive a differential energy spectrum. The resulting spectrum is shown
in \fref{spectrum} together with the spectrum published by the MAGIC
collaboration \cite{crabmagic}.
\begin{figure}[t]
   \centering
   {%
   \begin{overpic}[width=0.9\textwidth]{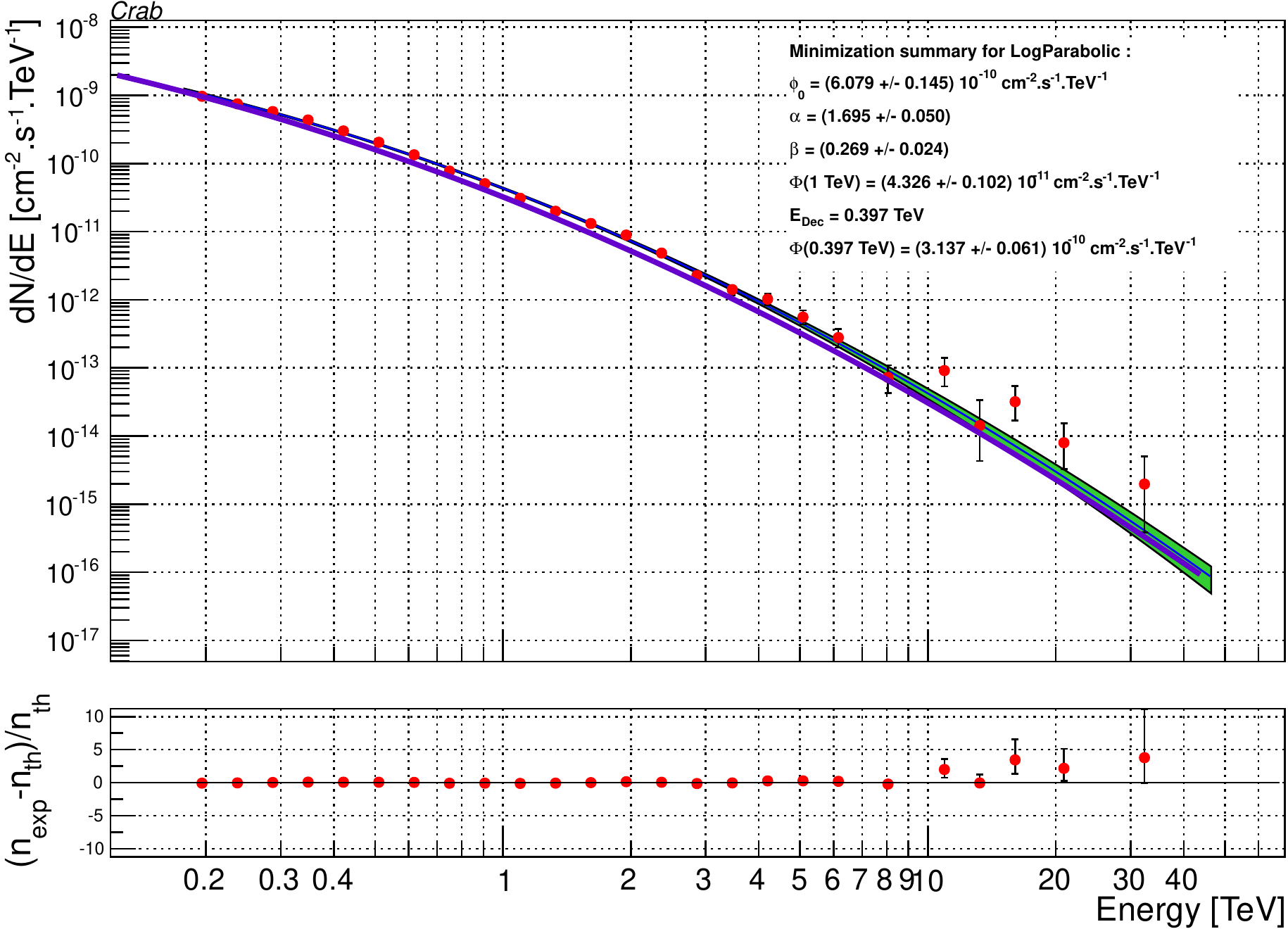}
     \put(20,25){\huge \textcolor{red}{\hess Preliminary}}
   \end{overpic} }
   \caption{Reconstructed differential energy spectrum of the Crab Nebula as a
           function of the true energy. The purple line represents the Crab
           spectrum published by the MAGIC collaboration \cite{crabmagic}.}
   \label{fig:spectrum}
\end{figure}
The spectral points are well-fit by a log-parabolic function. The fit parameters
are given in the inset. The reference energy is \SI{0.27}{TeV}. The spectrum is compatible with the result obtained by
the MAGIC collaboration within systematic uncertainties.

\section{Summary and Outlook}
A new method of reconstructing the main properties of \vhe γ rays has been
presented. The performance of the different parts of this reconstruction chain
is competitive to other reconstruction concepts as discussed above. A big
advantage of this chain is the speed with which events can be analysed. On
average only \SI{0.8}{\second} of CPU time are required to analyse 5000 events,
compared to approximately \SI{20}{\second} in the case of ImPACT.
This makes the chain well-suited for an analysis of newly taken data on site in
Namibia shortly after observations have been performed, enabling the \hess
collaboration to check for flux variability in the observed sources. Based on this,
decisions can be made whether to continue observing certain sources during the
following nights.

It will always be important to have a monoscopic reconstruction chain available
with which the entire energy range accessible to \hess can be analysed, for example for
split observation campaigns with only \ct{5} observing a source. Still it is
desirable to combine both monoscopic and stereoscopic reconstruction algorithms
into one \textit{combined} analysis, making it possible to decide on an
event-by-event basis whether a monoscopic or a stereoscopic reconstruction
algorithm should be used. Such an improvement will be provided in the near
future.


\begin{thebibliography}{99}
\bibitem{tevcat}
Online Gamma-Ray Catalog, \href{http://tevcat.uchicago.edu/}{http://tevcat.uchicago.edu/}
\bibitem{hess}
The \hess Collaboration, \href{http://www.mpi-hd.mpg.de/hfm/HESS/}{http://www.mpi-hd.mpg.de/hfm/HESS/}
\bibitem{fermi}
The Fermi Large Area Telescope, \href{http://www-glast.stanford.edu/}{http://www-glast.stanford.edu/}
\bibitem{hillasparams}
A. M. Hillas. Cerenkov light images of EAS produced by primary gamma. International Cosmic Ray Conference. 3:445–448, 1985
\bibitem{tmva}
A. Hoecker et al. TMVA: Toolkit for Multivariate Data Analysis. PoS, ACAT:040, 2007
\bibitem{impactarxiv}
R.D. Parsons and J. Hinton. A Monte Carlo Template based analysis for Air-Cherenkov Arrays, arXiv:1403.2993 [astro-ph.IM]
\bibitem{impacticrc}
R.D. Parsons, M. Gajdus and T. Murach. HESS II Data Analysis with ImPACT, Proceedings of the 34th International Cosmic Ray Conference, 2015
\bibitem{hesssens}
M. Holler et al. Photon Reconstruction for H.E.S.S. Using a Semi-Analytical Shower Model, Proceedings of the 34th International Cosmic Ray Conference, 2015
\bibitem{modelpp}
M. de Naurois and L. Rolland. A high performance likelihood reconstruction of γ-rays for imaging atmospheric Cherenkov telescopes, Astroparticle Physics, 32:231-252, 2009
\bibitem{bgpaper}
D. Berge, S. Funk and J. Hinton. Background Modelling in Very-High-Energy gamma-ray Astronomy, arXiv:astro-ph/0610959
\bibitem{lima}
T.-P. Li and Y.-Q. Ma. Analysis methods for results in gamma-ray astronomy. Astrophysical Journal, 272:317–324, September 1983. doi: 10.1086/161295.
\bibitem{crabpos}
A.~P. Lobanov, D. Horns and T.~W.~B. Muxlow. VLBI imaging of a flare in the Crab nebula: more than just a spot, \href{http://cdsbib.u-strasbg.fr/cgi-bin/cdsbib?2011A\&A...533A..10L}{http://cdsbib.u-strasbg.fr/cgi-bin/cdsbib?2011A\&A...533A..10L}
\bibitem{crabmagic}
R. Zanin et al., arXiv:1406.6892 [astro-ph.HE]

\end{thebibliography}
\end{document}